\documentclass[useAMS,usenatbib]{mn2e}

\def\aap{AA}
\def\apjl{ApJL}
\def\apjs{ApJS}
\def\mnras{MNRAS}
\def\apj{ApJ}
\def\aj{AJ}
\def\pasp{PASP}

\usepackage{graphicx}
\usepackage{float}
\usepackage{bm} 
\usepackage{amssymb}
\usepackage{amsmath} 
\usepackage[export]{adjustbox}
\usepackage{mathrsfs} 
\usepackage{mathtools}

\title[]{Ultra-diffuse galaxies: the high-spin tail of the abundant dwarf galaxy population}
\author[N. C. Amorisco \& A. Loeb]{N. C. Amorisco$^{1,2}$\thanks{E-mail:
nicola.amorisco@cfa.harvard.edu}, A. Loeb$^{1}$ \\
$^{1}$Institute for Theory and Computation, Harvard-Smithsonian Center for Astrophysics, 60 Garden St., MS-51, Cambridge, MA 02138, USA\\
$^{2}$Max Planck Institute for Astrophysics, Karl-Schwarzschild-Strasse 1, D-85740 Garching, Germany}

\begin{document}



\maketitle

\label{firstpage}

\begin{abstract}
Recent observations have revealed the existence of an abundant population of faint, low surface brightness (SB)
galaxies, which appear to be numerous and ubiquitous in nearby galaxy clusters, including the Virgo, Coma and Fornax clusters. 
With median stellar masses of dwarf galaxies, these ultra-diffuse galaxies (UDGs) have unexpectedly large 
sizes, corresponding to a mean SB of $24\lesssim\langle\mu_e\rangle_r\ {\rm mag}^{-1} {\rm arcsec}^2\lesssim27$ 
within the effective radius.
We show that the UDG population represents the tail of galaxies formed in dwarf-sized haloes with higher-than-average 
angular momentum. By adopting the standard model of disk formation -- in which the size of galaxies is set by the spin of the halo --
we recover both the abundance of UDGs as a function of the host cluster mass and the distribution of sizes within the UDG population.
According to this model, UDGs are not failed $L_*$ galaxies, but genuine dwarfs, and their low SB is not
uniquely connected to the harsh cluster environment. We therefore expect a correspondingly abundant 
population of UDGs in the field, with likely different morphologies and colours. 
\end{abstract}

\begin{keywords}
galaxies: dwarf --- galaxies: structure --- galaxies: formation --- galaxies: haloes  ---  galaxies: clusters \end{keywords}

\section{Introduction}

The existence of faint and extended galaxies in both field and clusters is certainly 
not a recent discovery \citep[e.g., ][]{Imp88, Both91, Tur93, JD97a, Cald06}. However, deep 
and wide field observations of nearby galaxy clusters have only recently first highlighted that this population is both
ubiquitous in clusters and perhaps surprisingly numerous. Using the Dragonfly array \citep{AvD14}, \citet{vanD15a} (vD15) first 
identified 47 extended, roundish, quiescent galaxies, with central surface brightness (SB) of $\mu_{g,0}=24-26$ mag arcsec$^{-2}$, 
in imaging data of the Coma cluster. 
Their projected spatial clustering and the inability of Hubble ACS observations to resolve them into stars suggested that 
these are actual cluster members. If so, at a distance of approximately 100 Mpc, these galaxies would have 
a median stellar mass of only $\approx6\times10^7 M_{\odot}$, but the sizes of $L_*$ galaxies, $r_e=1.5-4.6$ kpc. 
In fact, membership to Coma has been confirmed for one of them with Keck spectroscopy
\citep{vanD15b}, ascertaining the unusual properties of such ultra-diffuse galaxies (UDGs).

Since then, UDGs have been identified in significant numbers also in the Virgo cluster \citep{Mih15},
the Fornax cluster \citep{Mun15}, in 8 other clusters at redshift $z=0.044-0.063$ \citep[][vdB16]{vdB16}, and the number of detections in 
Coma have increased by at least an order of magnitude \citep[][K15]{Koda15}. 
However, the nature of their properties has remained elusive so far.

UDGs do not seem to show any clear evidence for tidal stripping. Also, their colours and 
spatial distributions within the host clusters imply that many of them have not just been accreted,
and that they are capable of surviving the strong cluster tides at pericenter. vD15 soon realised that they have to be
dark matter dominated systems. 
In opposition to classical low SB galaxies \citep[e.g.,][]{vdH93, McG94, Sch11}, UDGs appear quenched, and populate the red sequence.
This has brought vD15 and K15 to suggest that their existence could be strictly linked with the cluster environment,
and possibly that UDGs might represent `failed' $L_*$ galaxies, prematurely quenched at infall by gas removal.  
 
Measurements of the internal kinematics of UDGs would constrain their true masses 
and discriminate between different formation pathways, however this is challenging due to their low SB.
Very recently, \citet{Bea16} have taken advantage of the unusually abundant 
globular cluster system of VCC 1287, a UDG in the Virgo cluster, and, based on globular cluster spectroscopy inferred
a virial mass of only $(8 \pm 4)\times 10^{10} M_\odot$. This shows that VCC 1287 is a genuine dwarf galaxy despite its 
effective radius of $2.4$ kpc, disfavouring the possibility of a `failed' $L_*$ galaxy. Interestingly, similar conclusions 
can be drawn from the hydrodynamical simulations of \citet{YB15}. 
These authors show that larger-than-average disky dwarfs infalling onto a cluster are quickly quenched 
by ram pressure stripping, and reproduce the sizes and colours of many observed UDGs.

In this paper, we show that the properties and abundance of UDGs in clusters are largely compatible
with the classical model of disk formation \citep[e.g.,][]{FE80, JD97b, MMW98, AD07}. As
invoked for classical low SB galaxies \citep{JD97a, JD97b, MMW98,YB15},
galaxies formed in haloes with a higher initial spin result in 
larger disks, and therefore, low SB. Here we show that the UDG population discovered so far
represents the tail of those dwarf galaxies that formed in haloes with higher-than-average initial
angular momentum. Section~2 illustrates our simple model and tests it against the
population of normal, high SB galaxies. Section~3 compares the prediction of
our fiducial model with the observed properties and abundances of UDGs. Section~4 discusses results
and draws conclusions.

\begin{figure}
\centering
\includegraphics[width=.9\columnwidth]{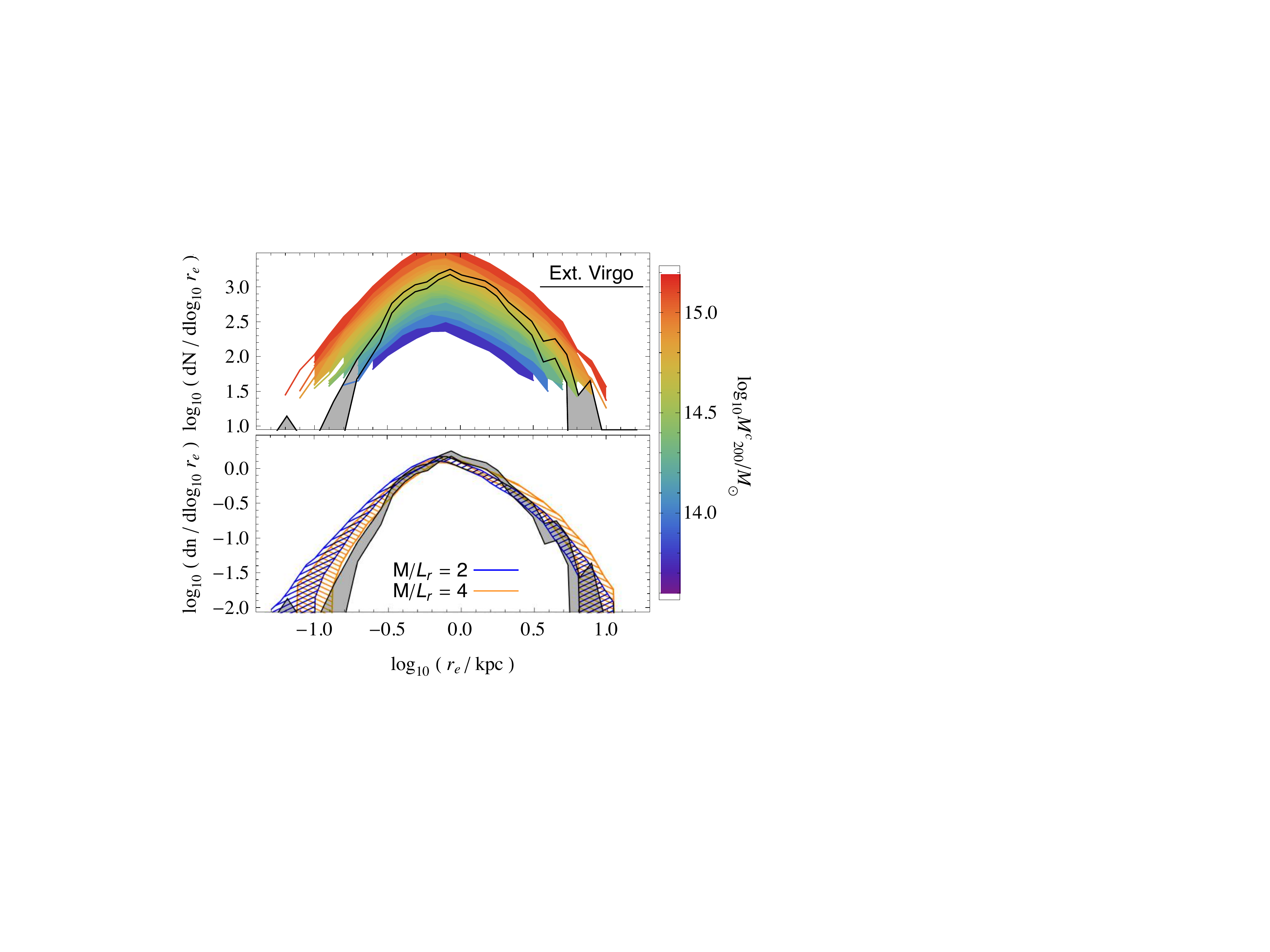}
\caption{Calibration of our fiducial model against the high SB galaxy population. In both the upper and lower panels,
black lines and grey shaded region identify the size distribution of members of the Virgo cluster \citep{Kim14}. The upper and 
lower panels compare this with cluster populations obtained from our fiducial model, in terms of absolute galaxy numbers, $dN$, and
relative distributions, $dn=dN/N_{be}$ ($N_{be}$ being the total number of galaxies within the cluster that comply with the cuts
that define the Virgo members). Models in the upper panel are colour-coded according to the virial mass of the cluster, $M^{\rm c}_{200}$,
the hatched regions in the lower panel illustrate the effect of different mass-to-light ratios.}
\label{virgocalibr}
\end{figure}

\section{$\Lambda$CDM galaxy populations}

We are interested in predicting abundances of galaxies defined by a set of cuts in
absolute magnitude, SB and/or galaxy size, within galaxy clusters with mass $M^{\rm c}_{200}$.
We adopt a very simple model, in which
\begin{itemize}
\item{The full subhalo population of the cluster is obtained from {Monte Carlo generated merger histories, 
with the spectra of both virial masses and accretion redshifts} calibrated on the Millenium simulation \citep{Fak10};}
\item{Each of these haloes contains a galaxy with a stellar mass $M_*$, fixed by a standard abundance matching relation \citep[e.g.,][]{PB13, GK14};}
\item{The effective radius of each galaxy is calculated according to the spin parameter of its halo \citep{MMW98};}
\item{Galaxies are considered destroyed by $z=0$ if not compact enough to survive the cluster tidal gravitational field.}
\end{itemize}

We assume that the distribution of halo spin parameters
\begin{equation}
\lambda\equiv {{J |E|^{1/2}}{G^{-1} M^{-5/2}}}
\end{equation}
is log-normal, with a mean value of $\langle \ln\lambda \rangle=0.05$ and $\sigma(\ln\lambda)=0.52$,
independent of redshift and halo mass \citep[e.g.,][]{JB01, Vit02, AM07, Kim15}.  
The concentration of halos follows the redshift dependent mass-concentration relation compiled by \citet{LG08}, 
with a scatter of approximately $0.3$ dex \citep[e.g.,][]{AL14}. We adopt the stellar-halo mass relation $M_*(M_{200})$ 
proposed by \citet{GK14}, independently of redshift, with a scatter of $0.3$ dex \citep{PB13,NM16}.  
Also, deviations in the properties of each galaxy from the mean relations are assumed to be uncorrelated.

We calculate all effective radii using the model of \citet{MMW98}, with $m_d=j_d=0.05$, i.e. {the specific angular momentum of disks reflects that of their halos, as also supported by observations of star forming galaxies at intermediate redshifts \citep{AB15}}. We do not distinguish galaxies according to their morphology, implicitly assuming that the initial spin sets the size of stellar spheroids as it does for disks 
\citep{SW03,AK13}. We use $r_e\equiv R_h$, where $R_h$ is the nominal disk scale-length predicted by the model, 
as a function of halo mass, concentration, spin parameter, and formation redshift $z_{\rm f}$ of each galaxy. 
We calibrate this on average Milky Way-like halos, for which the model predicts $R_h=3.6$ kpc \citep[as from][]{JB13}
for $z_{\rm f}=z_{\rm min}=0.7$. Therefore, we take the formation redshift to be the largest between $z_{\rm min}$
and the infall redshift of each halo onto the cluster.

Our final cluster populations only include galaxies that survive the cluster tides according to the Roche criterion:
\begin{equation}
{{M^{\rm g}(<2 r_e)}\over{M^{\rm c}(< r_{\rm peri})}} > 3 \left({{2 r_e}\over{r_{\rm peri}}}\right)^3 \ ,
\label{tides}
\end{equation}
where $M^{\rm g}(<2 r_e)$ is the galaxy mass within twice its effective radius, and $M^{\rm c}(< r_{\rm peri})$ is the cluster mass
within the orbital pericenter of the galaxy. This is calculated at the accretion redshift, using $r_{\rm peri} = R^{\rm c}_{200}/2 c$, 
where $R^{\rm c}_{200}$ is the virial radius of the cluster at that time and $c$ its concentration, so that the resulting $r_{\rm peri}$ 
is representative of the common cosmological accretion process \citep[e.g.,][]{AB05,AW11,LJ15}.

We first test the performance of this model on the cluster population of normal, high SB galaxies.
We use the Extended Virgo Cluster Catalog \citep{Kim14}, which collects properties of the 1028 spectroscopic members 
that comply with the observational cuts
\begin{equation}
\left\{ \begin{array}{l}
M_r < -13.4\ {\rm mag}\\
\langle\mu_e\rangle_r < 24.5\ {\rm mag\ arcsec}^{-2}
\end{array}\right. \ ,
\label{virgocut}
\end{equation}
%
where we indicate the mean SB within the effective radius with $\langle\mu_e\rangle$.
In the following, we refer to this population with the term `bright end', so that its abundance in any given cluster is $N_{be}$. 
The size distribution of the Virgo galaxies is displayed in Figure~1 using a grey shaded area, and is compared with the predictions of our 
simple model, that we have used to produce 300 mock clusters, with virial masses $M^{\rm c}_{200}$ in the 
interval $13.6<\log_{10}(M^{\rm c}_{200}/M_{\odot})<15.3$. 
The upper panel shows the abundance of the bright end population as a function
of the cluster mass. $N_{be}$ depends mildly on the adopted value of the
mass-to-light ratio of the population, with $N_{be}=10^3$ corresponding to a virial mass of $\log_{10}(M^{\rm c}_{200}/M_{\odot}) =(14.8\pm0.2)$
for galaxies with $M/L_r=(3\pm1) M_\odot/L_\odot$, in good agreement with the mass of Virgo within the wide area covered by the Extended Virgo 
Cluster Catalog \citep[e.g.,][]{Fou01,PC04}. The lower panel of Fig.~1 compares the relative distribution of galaxy sizes within the bright end 
population, which matches well the measured bell-shape.

\begin{figure}
\centering
\includegraphics[width=.9\columnwidth]{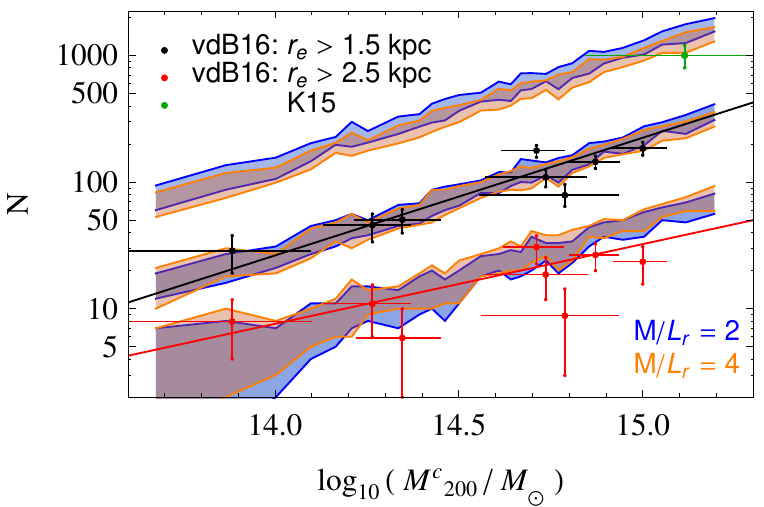}
\caption{Comparison of predicted UDG abundances $N$ with available measurements, as a function of the cluster virial mass. 
Data points and fitting lines display measurements from vdB16 and K15. Different colours refer to different defining criteria for the UDG population. 
Shaded regions extend between the 10 and 90\% quantiles for the corresponding populations in our model. 
Different colours indicate different mass-to-light ratios. }
\label{udgsnumb}
\end{figure}

\section{The UDG population}

The recent work of vdB16 is very useful to test our model, as it presents quantitative criteria 
defining the observed UDG samples. We apply the same cuts to our cluster galaxy populations and compare 
the results. First, we address the abundances $N$ of UDGs with the following set of properties
\begin{equation}
\left\{ \begin{array}{l}
M_r < -13.8\ {\rm mag}\\
24<\langle\mu_e\rangle_r\ {\rm mag}^{-1} {\rm arcsec}^2< 26.5\\
r_e > 1.5\ {\rm kpc}
\end{array}\right. \ ,
\label{vdbcut}
\end{equation}
as used by vdB16 to measure UDG abundances 
in 8 clusters at redshift $z\approx0.05$. Black points and the black guiding line
display such measurements in Figure~2. The underlying shaded regions cover the area between the 10\% and the 90\% quantiles
as predicted by our model for the same abundances, for two different values of the mass-to-light ratio. 
The quantitative agreement between model predictions and measurements is very good. 
Red data points and fitting line are obtained from the same observational dataset, but only include the largest
UDGs in the sample, with a more extreme cut of $r_e > 2.5\ {\rm kpc}$. As for the full UDG population, 
this tail is also quite well reproduced by our model. 
Finally, Fig.~2 displays, with a green point, the measurement performed by K15 for the abundance of UDGs in the Coma cluster.
They adopt the definition
\begin{equation}
\left\{ \begin{array}{l}
M_r < -12\ {\rm mag}\\
24.3<\langle\mu_e\rangle_r\ {\rm mag}^{-1} {\rm arcsec}^2< 27.3\\
r_e > 0.8\ {\rm kpc}
\end{array}\right. \ ,
\label{kodacut}
\end{equation}
which samples both fainter and less diffuse systems, therefore reaching an abundance of approximately 1000 UDGs in the cluster.
Applying these cuts to our galaxy populations produces similar abundances, as shown by the shaded regions. 
As the stellar M/L ratio is only weakly variable, we uniformly adopt $M/L_r=3\ M_\odot/L_\odot$ in the following.

\begin{figure}
\centering
\includegraphics[width=.825\columnwidth]{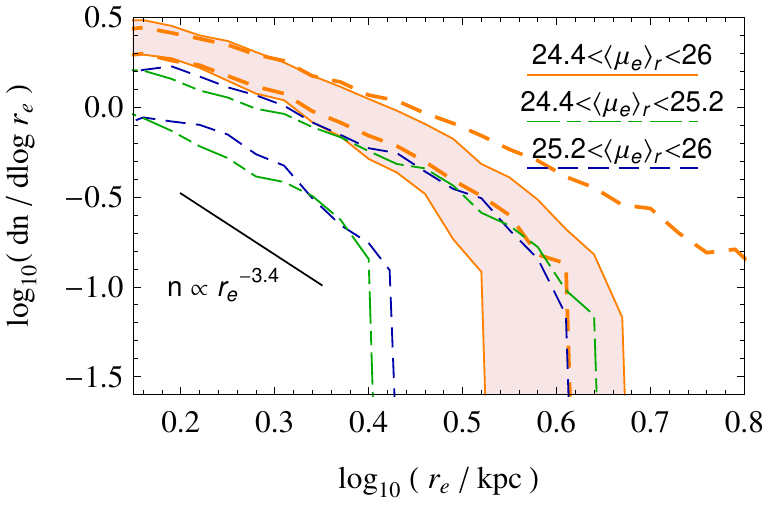}
\caption{The UDG size distribution. 
The shaded region shows the size distribution of cluster UDGs, defined as in vdB16, which agrees well with 
the measured slope $n\propto r_e^{-3.4}$, illustrated by the black guiding line {(field UDGs are displayed with an orange dashed profile)}.
The green and blue dashed styles display the size distribution of two subsets of the same UDG population, defined according
to SB as in the legend. As measured by vdB16, UDGs are approximately uniform in SB. }
\label{udgsnumb}
\end{figure}

vdB16 has measured the size distribution within the UDG population, defined adopting the cuts of eqn~(\ref{vdbcut}), 
but with a more restrictive SB constraint, $24.4<\langle\mu_e\rangle_r\ {\rm mag}^{-1} {\rm arcsec}^2< 26$.
They find a steep decline with size, well fit by a power-law relation $n\propto r_e^{-3.4}$. Furthermore, they find that the 
UDGs are approximately equally split between two equal-size bins in SB. Figure~3 shows the corresponding predictions 
from our model. We recover both a compatible decline in the size distribution, and the approximate uniformity in SB
of the UDGs defined this way. {The apparent truncation in the size distribution of our cluster UDG population 
is due to the criterion~(2), which forces all infalling UDGs on the same quite radial orbit; such truncation is absent in the field population.}

\begin{figure*}
\centering
\includegraphics[width=.825\textwidth]{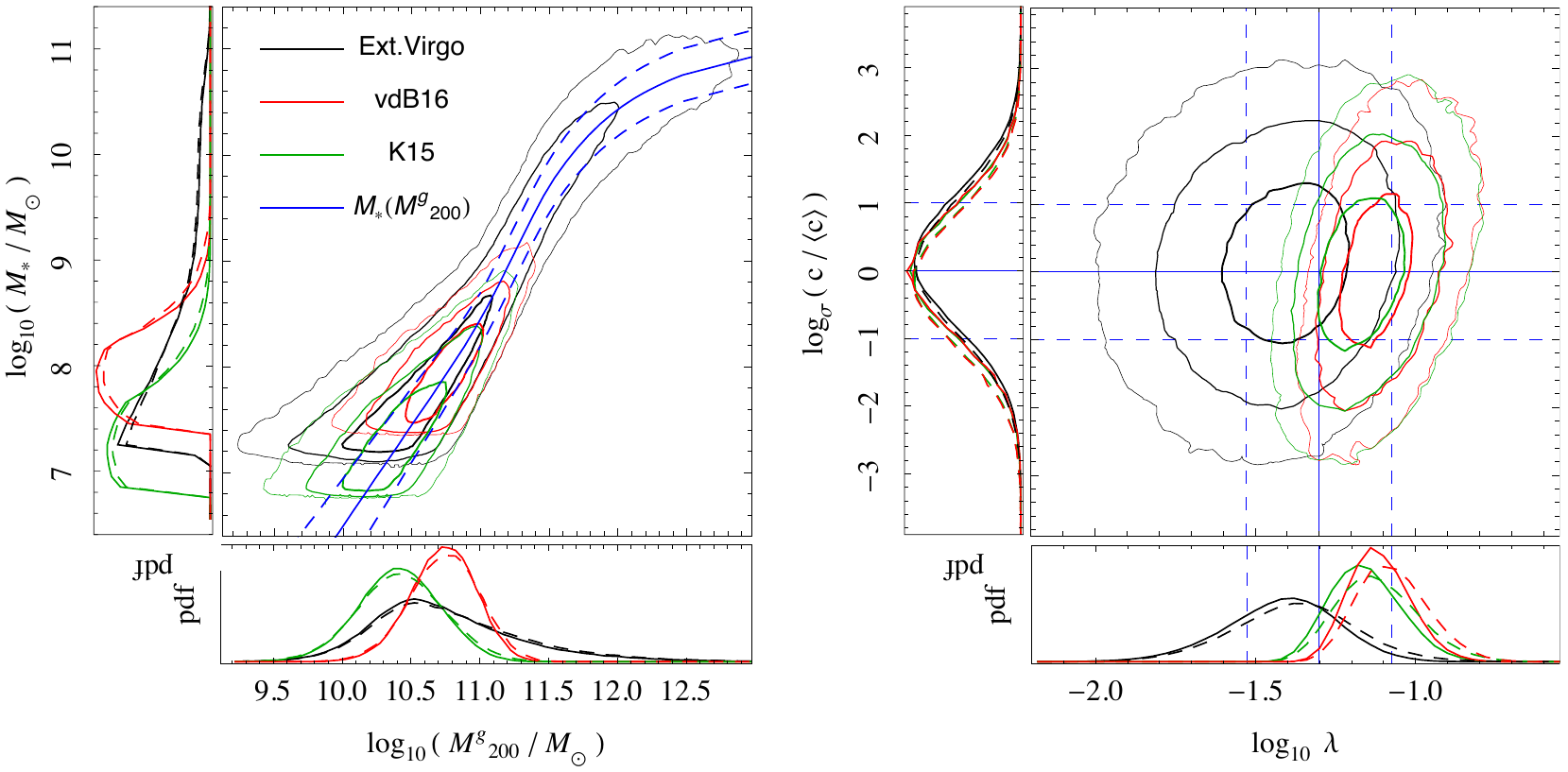}
\caption{The left and right panels illustrate the location of the UDGs in the plane $(M_{200}, M_*)$
and in the plane of halo spin against normalised deviation in the halo concentration. Contours represent 50, 90 and 99\% of galaxies. 
Blue lines show the adopted abundance matching relation in the left panel and the scatter in both spin and concentration  
distributions in the right panel. The smaller panels show the one-dimensional (projected) distributions. 
Full and dashed lines identify the cluster and field populations.}
\label{udgsnumb}
\end{figure*}
 
\subsection{Physical properties}

Figure~4 puts the properties of UDGs into context. The location of the different galaxy populations (\ref{virgocut}, \ref{vdbcut}, \ref{kodacut})
is shown in the plane of stellar-halo mass and in the plane of spin parameter against deviation from the mean halo concentration.
Contours of different thickness include 50, 90 and 99\% of the full populations. We find that UDGs are not failed $L_*$ galaxies, but 
genuine dwarf galaxies, with virial masses that, depending on the adopted observational cut, range in the interval $10^{10}-10^{11} M_{\odot}$.
{These follow from the adopted abundance matching relation, which however remains quite uncertain at the low stellar masses relevant to the UDGs}, $10^7 - 10^{8.5} M_\odot$ \citep[e.g.,][]{GK16}. UDGs do not seem to diverge from the mean stellar-to-halo mass relation, 
at least not more than the bright end population is oppositely biased towards systems that are more luminous than average. 
Their concentrations are also average, with a bias towards less concentrated haloes that is similar but opposite in sign to the bias towards 
higher concentrations seen in the bright end population. Similarly, we identify {only limited differences in the distribution
of accretion times, with a tendency for UDGs accreted at higher redshifts to be more easily shredded by the cluster tides,
due to the redshift dependence of the halo density contrast \citep[e.g.,][]{NA15}}. What clearly distinguishes UDGs and the bright end 
galaxies are their spin distributions.
UDGs are concentrated in the high-spin tail of the halo population, centered approximately one sigma 
away from the mean. The bright galaxies are also biased, but by a much smaller amount and, of course, in the opposite direction. 

Dashed profiles in the one-dimensional insets show the distribution of the different populations when the galaxies excised according to 
requirement~(\ref{tides}) are reincorporated. Under the reasonable assumption that the population of haloes infalling onto clusters
have a similar mass distribution compared to those that remain isolated, these profiles can be considered representative of the corresponding 
galaxy populations in a field environment. Field and cluster populations do not display significant differences in their mean physical properties, 
apart from the 
tendency of field populations to shift slightly towards higher spin parameters (and therefore larger sizes), illustrating the fact that,
according to our model, the most diffuse UDGs are indeed shredded by tides when infalling onto clusters. 

\section{Discussion }

We have shown that the unexpectedly large size and low SB of UDGs are reproduced naturally
within the same standard framework that describes the properties of normal high SB galaxies. 
The combination of a $\Lambda$CDM population of haloes and a classical abundance matching relation is 
sufficient to reproduce the significant numbers of UDGs observed in galaxy clusters, as a result of the 
extremely abundant population of dwarf galaxies and the sharp decline in the $M_*(M_{200})$ relation at low masses.
UDGs are not a different or peculiar new population of galaxies, but rather the natural extension of normal dwarfs
into the low SB regime, characterised by high-spin haloes. 

According to our simple criterion for survival against the tides, several UDGs are indeed not compact enough, 
confirming that a cluster environment is a harsh one for these tenuous dwarfs
and suggesting that a fraction of the observed UDGs may in fact be on the verge of tidal disruption. 
Indeed, as noticed by K15, the morphological properties of the observed UDGs are incompatible 
with them being a population of disks, and their transformation from very extended disks to thin 
spheroids may well be caused by the interaction with the cluster, 
through ram pressure stripping and tidal stirring \citep[e.g.,][]{LM01,SK11}. 

\begin{figure}
\centering
\includegraphics[width=.825\columnwidth]{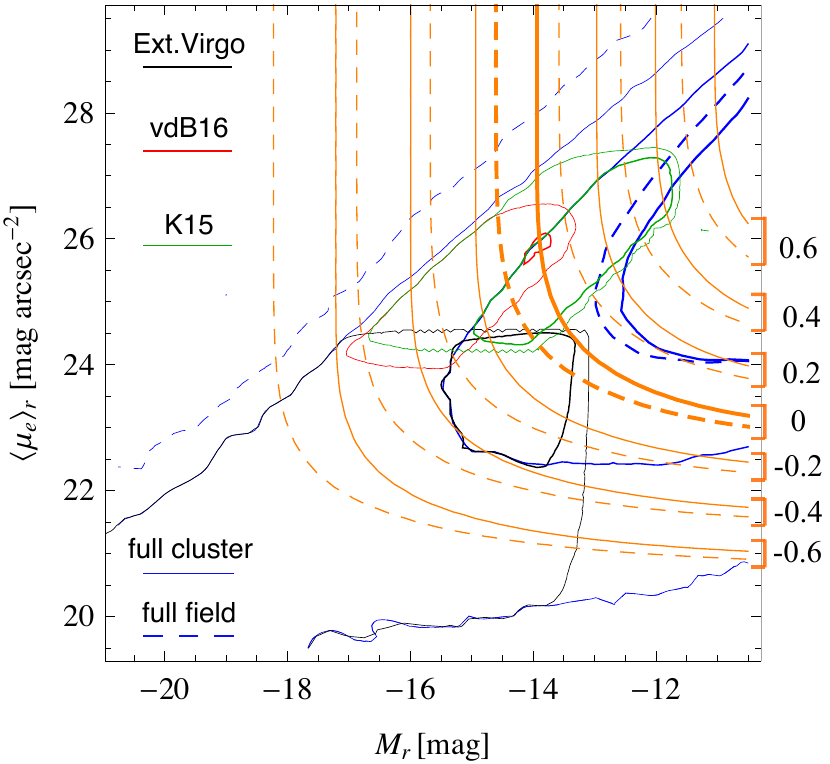}
\caption{The number density of galaxies in the plane $(M_r, \langle\mu_e\rangle_r)$, relative to the number in the bright 
cluster population, $N_{be}$. Colour-coding is as in Fig.~4, with blue contours representing the full galaxy population. 
The contour sets identify the density values $\{10^{-2.2},10^{-1.2},10^{-0.6}\}$. 
Orange contours show the quantity $\log_{10}\left[N(M_r,\langle\mu_e\rangle_r)/N_{be}\right]$, where $N(M_r,\langle\mu_e\rangle_r)$ 
is the total number of galaxies brighter than $M_r$ and with SB higher than $\langle\mu_e\rangle_r$. Full and dashed profiles 
(in both blue and orange) refer respectively to the cluster and field populations.}
\label{udgsnumb}
\end{figure}

However, the cluster itself is not strictly necessary for the formation of dwarf galaxies with such a low SB. 
Under the assumption that the spin distribution is not strongly dependent on environment {\citep[e.g.,][]{AM07,Kim15}}
and that these extended disks are capable of forming stars in a similar way when in isolation, our model
suggests that an abundant tail of extended galaxies should be ubiquitous in both clusters and in the field. {In fact, 
\citet{MD16} have recently discovered a UDG, DGSAT I, in the outskirts of the Pisces-Perseus supercluster,  
in a considerably less dense environment than the other known UDGs. DGSAT I displays an out-of-center, 
bluer over-density, compatible with a clump of recently formed stars, and suggesting that the isolated counterparts
of cluster UDGs may have more clearly disky morphologies, and not appear as red and quenched.
After this manuscript was submitted \citet{RT16} reported the detection of a numerous population of UDGs
in the large scale structure surrounding the cluster Abel 168, confirming that UDGs can indeed form outside clusters. 
In support of our model, the same authors also find that the spatial distribution of UDGs is almost identical to that of 
normal dwarf galaxies, while it is significantly different from the one of $L_*$ galaxies.}

Finally, to facilitate comparisons with future observations, Figure~5 provides predictions for the abundance 
of galaxy populations as a function of the defining cuts, 
relative to the cluster population in the bright end, $N_{be}$. Orange lines display contours of the quantity 
\begin{equation}
\log_{10} n_{(M_r, \langle\mu_e\rangle_r)} = \log{N_{(M_r, \langle\mu_e\rangle_r)}}-\log{N_{(-13.4, 24.5)}} 
\label{relabu}
\end{equation}
where $N_{(M_r,\langle\mu_e\rangle_r)}$ is the number of galaxies brighter than $M_r$ and with SB higher 
than $\langle\mu_e\rangle_r$, so that $N_{(-13.4, 24.5)}=N_{be}$. Full and dashed lines refer to the cluster and,
even more numerous, field populations. 

Our simple model does not directly address the detailed evolutionary pathways of UDGs,
how such thin galaxies can indeed form stars, or whether they are made even more dark matter 
dominated when in a cluster environment with respect to the field. Follow up spectroscopic studies
and perhaps the observation of field UDGs will allow better constraints on these issues.

\section*{Acknowledgments}
This work was supported in part by NSF grant AST-1312034 (for A.L.).



\end{document}